\documentclass[jpurnal=jpcl,showpacs,floatfix,nofootinbib,preprintnumbers,superscriptaddress,amsmath,amssymb]{achemso}

\usepackage[dvipsnames]{xcolor}
\usepackage{chemformula} 
\usepackage[T1]{fontenc} 

\usepackage{hyperref}
\usepackage{bm}
\usepackage{booktabs}
\usepackage[font=small,labelfont=bf]{caption}
\usepackage{soul}
\usepackage{graphicx}
\usepackage[version=3]{mhchem} 
\usepackage{multirow}

\usepackage{amsfonts,calrsfs}
\usepackage{amsmath}

\usepackage{amssymb}
 
\newcommand\myshade{85}
\colorlet{mylinkcolor}{violet}
\colorlet{mycitecolor}{NavyBlue}
\colorlet{myurlcolor}{Aquamarine}

\hypersetup{
  linkcolor  = mylinkcolor!\myshade!black,
  citecolor  = mycitecolor!\myshade!black,
  urlcolor   = myurlcolor!\myshade!black,
  colorlinks = true,
}

\SectionsOn
\SectionNumbersOn

\author{Andres Robles-Navarro}
\affiliation{Centre for Theoretical Chemistry and Physics, The New Zealand Institute for Advanced Study, Massey University Auckland, Private Bag 102904, 0745 Auckland, New Zealand}
\author{Shaun Cooper} 
\affiliation{School of Mathematical and Computational Sciences, Massey University Auckland, Private Bag 102904,
Auckland 0745, New Zealand}
\author{Odile R. Smits}
\affiliation{Centre for Theoretical Chemistry and Physics, The New Zealand Institute for Advanced Study, Massey University Auckland, Private Bag 102904, 0745 Auckland, New Zealand, and the School of Mathematics and Physics, University of Queensland, Brisbane QLD 4072, Australia}
\author{Peter Schwerdtfeger}
\affiliation{Centre for Theoretical Chemistry and Physics, The New Zealand Institute for Advanced Study, Massey University Auckland, Private Bag 102904, 0745 Auckland, New Zealand}
\email{peter.schwerdtfeger@gmail.com}

\title[Exploring the mechanism of phase transitions between the hexagonal close-packed and the cuboidal structures]{Exploring the mechanism of phase transitions between the hexagonal close-packed and the cuboidal structures}

\keywords{Diffusionless phase transitions, Cohesive energies, Hard-sphere model, Lennard-Jones potential, Lithium}

\begin{document}


\begin{abstract}
By introducing appropriate lattice parameters for a bi-lattice smoothly connecting the hexagonal close-packed (hcp) with the cuboidal structures, namely the body-centered (bcc) and the face centered cubic (fcc) lattices, we were able to map out the minimum energy path for a Burgers-Bain type of phase transition. We demonstrate that for three different models applied, i.e. the kissing hard-sphere model, the Lennard-Jones potential, and density functional theory for metallic lithium, the direct transition path is always from hcp to fcc with a separate path leading from fcc to bcc. This solves, at least for the models considered here, a long-standing controversy of whether or not fcc acts as an intermediate phase in martensitic type of phase transitions.
\end{abstract}

\vspace{5cm}
\begin{figure}[]
\centering
\textbf{TOC Graphic}\\
\vspace{1cm}
    \includegraphics[scale=1.5]{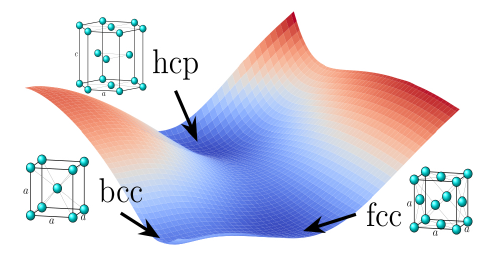}
\end{figure}

\maketitle

\newpage
\section*{}


 Solid-state phase transitions are notoriously difficult to model \cite{Liu2018}. While experimental methods can give quite detailed $(P,T,V,\mu,...)$ phase diagrams, the specific mechanism (dynamics) of a phase transition around the boundary line of different phases often eludes such experiments even for the simplest solids such as the rare gases. On the computational side, phase transition simulations are computationally very demanding and often plagued by the accuracy of the underlying model chosen for describing the intermolecular forces \cite{Verlet1969,CHANG2004}. Unlike in simple chemical reactions where the minimum energy path (MEP) from the reactants to the product can be obtained by certain algorithms \cite{Schlegel2003} and sophisticated electronic structure methods such as coupled-cluster theory \cite{Bartlett2007}, for the solid state it is often difficult to map out phase transition MEPs \cite{Carter2005}. Here one relies mostly on model potentials \cite{Travesset2014} or on computationally demanding density functional theory and sophisticated algorithms like machine learning to explore the potential energy hypersurface \cite{Kresse1998,Kresse2019,Kresse2021}.

Martensitic transformations are a subclass of phase transitions described by diffusionless displacive transformations caused by lattice deformations where atoms move in a coordinated, homogeneous, and crystallographically oriented manner relative to their neighbors along some MEP \cite{Grimvall2012,nishiyama2012,lobodyuk2014}. It controls for example the mechanical properties of many materials such as steel at elevated temperatures \cite{Venables1962}. Bain described such a transition from fcc to bcc (the so-called Bain path) as early as in 1924 on purely crystallographic grounds \cite{Bain1924}, and Burgers later in 1934 in a similar fashion for the hcp to bcc transformation (Burgers path) \cite{Burgers1934}. While the former only involves changes in the basic lattice parameters $a$ and $\gamma_{2}=c/a$ within a cuboidal lattice, the latter is more complicated and involves a transition from a hexagonal based bi-lattice to some cuboidal (cub) lattices such as bcc \cite{Grimvall2012,Cayron2015}. It remains, however, an open question whether there is a direct path from hcp to bcc or if the fcc lattice is an intermediate step in the Burgers hcp$\rightarrow$bcc transformation \cite{Olson1976,Bruinsma1985,Akahama2006,Lu_2014}. For a recent discussion on lattice instabilities see Grimvall et al  \cite{Grimvall2012}.

In order to map out the MEP between hcp and the cuboidal lattices, we developed a simple and intuitive lattice model which we apply to three different models: the hard-sphere model \cite{Baxter1968}, the (12,6)-Lennard-Jones potential \cite{Wales2024}, and to metallic lithium using density functional theory \cite{gross2013}. Our treatment involves a much larger parameter space than previously applied \cite{Straub1971,Li1999,Natarajan2019} and should therefore give a detailed insight into the combined Burgers-Bain type of phase transition.

\begin{figure}[htb!]
\centering
\includegraphics[scale=1.0]{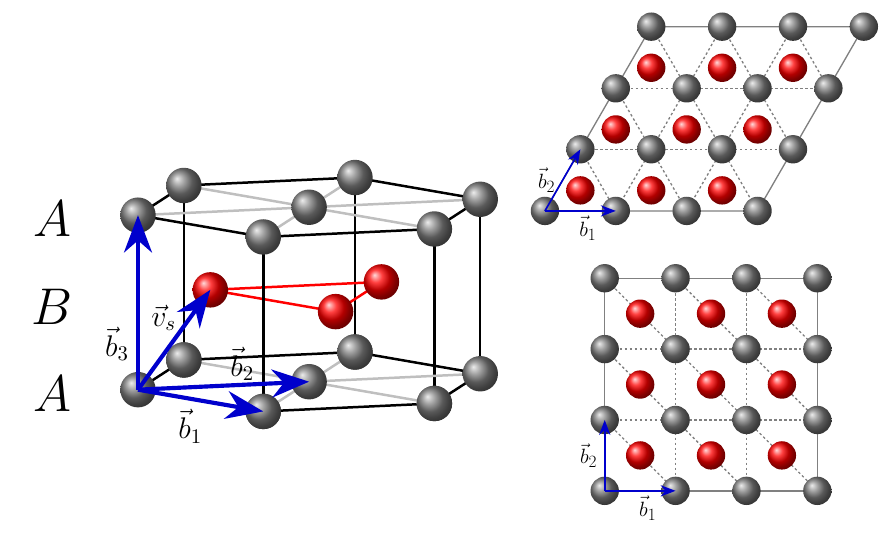}
\caption{Left: The (distorted) cuboidal (blue lines shown on the left) and the hcp structure both with a ABABAB... sequence (layers A in red and B in blue) in a hexagonal unit cell with corresponding basis vectors. Right: The transformation between the cuboidal lattices fcc or bcc (top) and the hexagonal closed packed bi-lattice (bottom). The red atoms indicate the middle layer in the unit cell.}
\label{fig:Burgers}
\end{figure}

 We define the generator matrix $B=\{\vec{b}_1,\vec{b}_2,\vec{b}_3\}$ containing the lattice vectors $\vec{b}_i$ and the shift vector $\vec{v_s}$ as follows
\begin{align}\label{eq:Bmatrix}
\nonumber
B&=a 
\begin{pmatrix}
1  & 0  &  0 \\
\frac{1}{2}\gamma_1(1-\alpha) &\frac{1}{2}\gamma_1\sqrt{(1+\alpha)(3-\alpha)} &  0 \\
0 & 0  &  \gamma_2
\end{pmatrix}\\
\vec{v}_s^\top&=\frac{a}{2}(\beta_1,\beta_2,\beta_3)
\end{align}
The $B$-matrix has been obtained from the conditions that (a) we have a linear transformation between the hexagonal and the cuboidal (cub) $B$-matrices defining the unit cell and characterized by the transformation parameter $\alpha$, (b) we have a bi-lattice introducing a middle layer with the atom in the unit cell situated at the Wyckoff position $\vec{v}_s$ as defined in eq.\eqref{eq:Bmatrix} and shown in Figure \ref{fig:Burgers}, (c) any distortion from the base lattice parameter $|\vec{b}_1|=|\vec{b}_2|=a$ is described by the lattice parameter $\gamma_1$ ($|\vec{b}_2|=\gamma_1a$), and (d) we define $\gamma_2=c/a$ as the ratio between the two lattice parameters in a Bravais lattice. We keep the angles between the vectors $\angle(\vec{b}_{3},\vec{b}_{1})$ and $\angle(\vec{b}_{3},\vec{b}_{2})$ at 90$^\circ$ reducing the problem from a complete 9 to a subset of 7 parameters for the lattices constants $p_i$, i.e. $\{p_i\}=\{\alpha,a,\gamma_1,\gamma_2,\beta_1,\beta_2,\beta_3\}$. 

The transformation from bcc/fcc to hcp is visualized in Figure \ref{fig:Burgers}. The parameters for the three lattices are summarized in Table \ref{tab:latticeparameters}. From the table it is clear that the lattice parameter $\alpha$ is responsible for the hcp$\rightarrow$cub (cub = fcc or bcc) transformation through the angle $\theta_{12}$ between the vectors $\vec{b}_{1}$ and $\vec{b}_{2}$, i.e. 
$\cos\theta_{12}=\frac{1}{2}(1-\alpha)$, and the parameter $\gamma_2$ for the fcc$\rightarrow$bcc transformation.\cite{Chen1988} This implies that we can describe the phase transition as a hypersurface of the cohesive energy, $E_\text{coh}(\alpha,\gamma_2)$ with the remaining lattice parameters being optimized.

\begin{center}
\begin{table}[hbtp!] 
\setlength{\tabcolsep}{8pt}
\begin{tabular}{ |l|r|r|r| } 
 \hline
 Parameter  & hcp                   & fcc       & bcc \\ 
 \hline 
$\alpha$    & $0$                 & $1$       & $1$ \\
$a$         & $1$                 & $1$       & $\frac{2}{\sqrt{3}}$ \\
$\gamma_1$  & $1$                 & $1$       & $1$ \\
$\gamma_2$  & $\sqrt{\frac{8}{3}}$  & $\sqrt{2}$  & $1$ \\
$\beta_1$   & $1$                 & $1$       & $1$ \\
$\beta_2$   & $\frac{1}{\sqrt{3}}$  & $1$       & $1$ \\
$\beta_3$   & $\sqrt{\frac{8}{3}}$  & $\sqrt{2}$  & $1$ \\
$\angle(\vec{b}_{1},\vec{b}_{2})$   & 60$^\circ$  & 90$^\circ$ & 90$^\circ$\\
$V$                                 & $\sqrt{2}$   & $\sqrt{2}$     & $\frac{8}{3\sqrt{3}}$ \\
$\rho$                              & $\frac{\pi}{3\sqrt{2}}$   & $\frac{\pi}{3\sqrt{2}}$     & $\frac{\pi\sqrt{3}}{8}$ \\
$N_\text{kiss}$      & 12         &12       & 8 \\   
\hline
\end{tabular}
\caption{Lattice parameters for kissing hard spheres of diameter 1 for the three lattices hcp, fcc and bcc. The volumes $V$, corresponding packing densities and kissing numbers $N_\text{kiss}$ are also given.}
\label{tab:latticeparameters}
\end{table}
\end{center}

We can derive the volume of the unit cell using the $B$-matrix from \eqref{eq:Bmatrix}
\begin{align}\label{eq:vol1}
V=\text{det}(B)= \frac{a^3\gamma_1\gamma_{2}}{2} \sqrt{(1+\alpha)(3-\alpha)}
\end{align}
Since $V>0$ we have the condition that $\alpha\in (-1,3)$.

The kissing hard-sphere (KHS) model with attractive inverse power potential, called the Sutherland potential,\cite{Sutherland1893} is defined by $\phi(r)=\infty$ for $r<1$ and $\phi(r)=-r^{-n}$ for $r\ge 1$ (in reduced units). This model maximizes the number of nearest neighbors with distance 1, i.e. the kissing (or coordination) number $N_\text{kiss}$. For the cohesive energy we obtain $E_\text{coh}=-\frac{1}{2}L_{\frac{n}{2}}(p_i)$, where $L_{\frac{n}{2}}(p_i)$ are 3D infinite lattice sums dependent on the exponent $n$ and the lattice parameters $\{p_i\}$ as defined in the generator matrix $B$. These are slowly converging sums which we treat efficiently to computer accuracy in terms of fast converging Bessel function expressions \cite{burrows-2020}. In the limit $n\rightarrow \infty$ we get $E_\text{coh}=-N_\text{kiss}/2$. The minimum energy conditions yields $a=1, \gamma_1=1$ and $\beta_1=1$ for the hcp$\rightarrow$cub transition, leading to the expressions for the lattice parameters and the volume $V$ derived from simple geometric arguments, 
\begin{align}
\gamma_2(\alpha)&=\sqrt{\frac{8+4\alpha(1-\alpha)}{4-(1-\alpha)^2}}\\
\beta_2(\alpha)&=\sqrt{\frac{4-(1-\alpha)(\alpha+3)}{4-(1-\alpha)^2}}\\
V(\alpha)&=\sqrt{2+\alpha(1-\alpha)}
\end{align}
$\gamma_2(\alpha)$ is a monotonically decreasing function in the interval $\alpha\in[0,1]$ changing from $\gamma_2(0)=\sqrt{\frac{8}{3}}$ to $\gamma_2(1)=\sqrt{2}$. Consequently, the KHS model directly connects the hcp with the fcc phase. The volume is at maximum at $\alpha=\frac{1}{2}$ with $V=\frac{3}{2}$, which we interpret as the transition state. Along the hcp$\rightarrow$fcc transition path the kissing number reduces to $N_\text{kiss}=10$ for the interval $\alpha\in(0,1)$. For the $r^{-6}$ attractive potential, for example, we locate the transition state at $\alpha^{\#}=0.5000623690$ with an activation energy (with respect to the hcp structure) of $\Delta E_\text{coh}^{\#}=0.7061966753$. The small deviation from the ideal $\alpha=\frac{1}{2}$ value comes from the fact that the hcp structure is slightly more stable compared to fcc by an energy difference of $\Delta E_\text{coh}=-4.881170486\times 10^{-4}$. 

If we proceed to the bcc phase the parameter $\gamma_2$ becomes now our reaction coordinate for the fcc$\rightarrow$bcc Bain path. Here the 8 atoms at the edges of the cuboidal cell move slightly outwards as the unit cell gets compressed along the $c$-axis and we have $a>r_\text{bc}=1$, where $r_\text{bc}$ is the distance from the sphere at the origin of the unit cell to the one in the middle layer. The kissing number changes now from 12 (fcc) to 8 along the path to bcc. We can easily derive that
\begin{equation}
a(\gamma_2)=\frac{2}{\sqrt{2+\gamma_2^2}}\quad \text{and} \quad
V(\gamma_2)=\frac{8\gamma_2}{(2+\gamma_2^2)^\frac{3}{2}}
\end{equation}
In this case the volume is monotonically increasing from fcc to bcc. For the $r^{-6}$ attractive potential, the bcc structure becomes a maximum at $\gamma_2=1.0$, with a difference to the fcc structure of $\Delta E_\text{coh}=1.100126588$, which lies energetically even above the transition state for the hcp$\rightarrow$fcc path. In this model the bcc phase is not stable towards distortion to the fcc structure \cite{burrows2021b}.

For a general $(n,m)$-LJ potential (in reduced units and $n>m$) \cite{Wales2024}
\begin{equation} \label{eq:VLJ}
\phi_{\textrm{LJ}}(r)=\frac{nm}{n-m} \;  \left[ \frac{1}{nr^n} - \frac{1}{mr^m} \right] 
\end{equation}
we get the cohesive energy as a function of the lattice constant $a$ for $m>3$ to avoid divergencies \cite{Schwerdtfeger-2006}
\begin{equation} \label{eq:cohLJhcp}
E_{\text{coh}}(p_i)= \frac{nm}{2(n-m)} \; \left[ \frac{L_\frac{n}{2}(p_i)}{na^n} - \frac{L_\frac{m}{2}(p_i)}{ma^m} \right]
\end{equation}
in terms of the lattice sums $L_\frac{n}{2}(p_i)$. As both lattice parameters connect the three phases hcp, fcc and bcc in a natural way, we map out the $E_{\text{coh}}(\alpha,\gamma_2)$ hypersurface to obtain a detailed insight into the phase transitions.
To save computational time we restrict three of the lattice parameters to their ideal values as shown in Table \ref{tab:latticeparameters}, i.e. $\gamma_1=1$, $\beta_3=\gamma_2$ and $\beta_1=1$. Test calculations with a few random points in the parameter space revealed that the former two restrictions can be imposed and the latter is justified as it does not change the overall topology of the hypersurface. All other lattice parameters were optimized by using a Newton-Raphson procedure \cite{hammerlin2012numerical}.

\begin{figure}[htb!]
    \centering
    \includegraphics[width=0.7\linewidth]{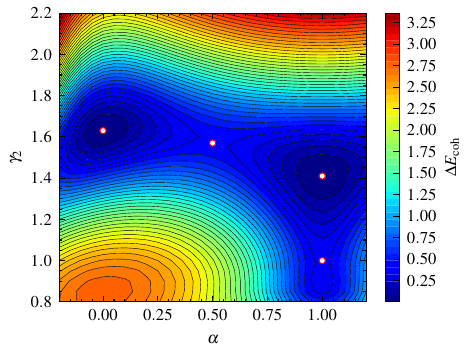}\\
    \includegraphics[width=0.7\linewidth]{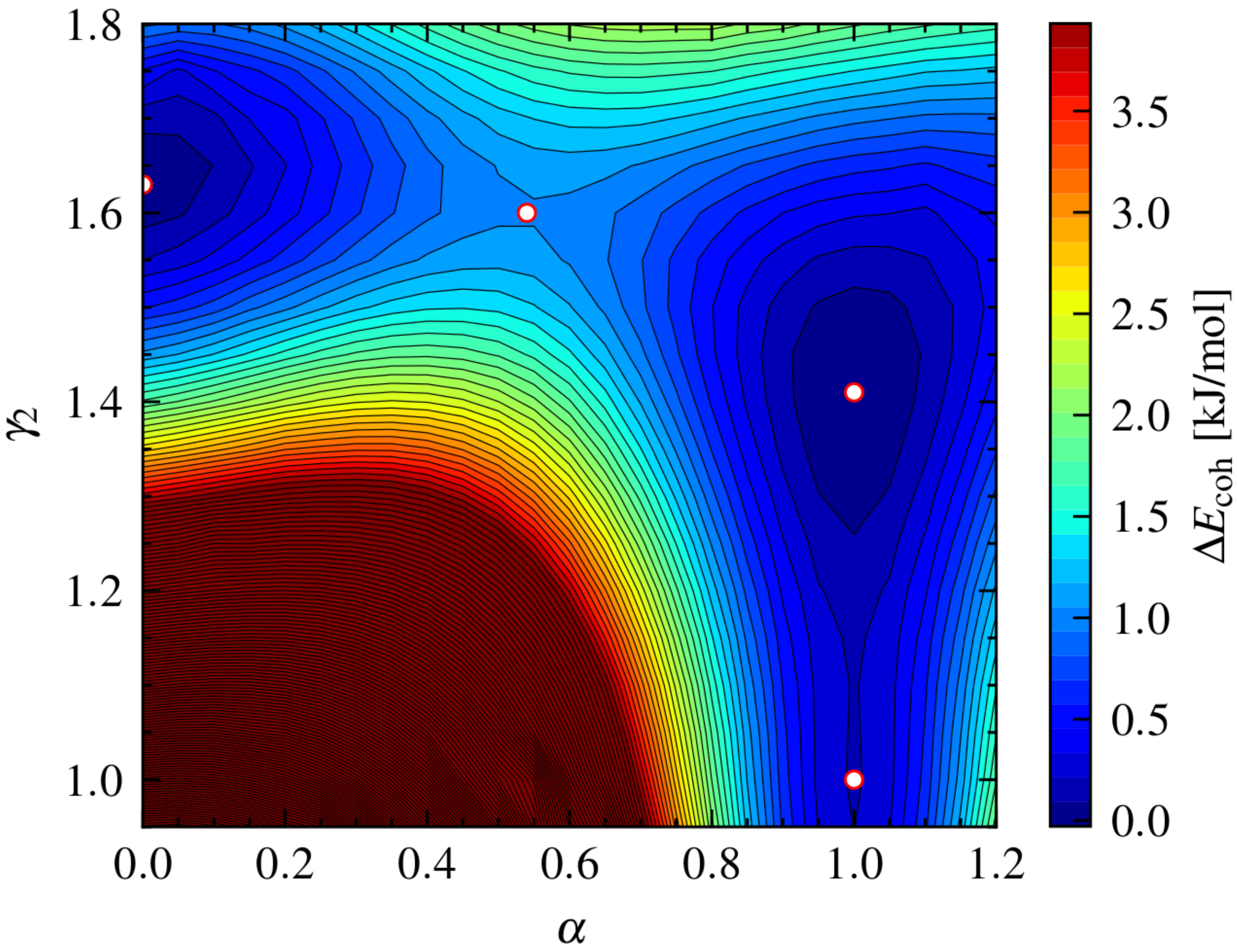}
    \caption{Cohesive energy hypersurface, $\Delta E_{\text{coh}}(\alpha,\gamma_{2})$, with all other lattice parameters optimized for (top) the $(12,6)$-LJ potential (in reduced units), and (bottom) the lithium metal (in kJ/mol) calculated using the PBE-D3 functional. The white circles represent the position of the hcp ($\alpha=0,  \gamma_2=\sqrt{8/3}$), hcp$\rightarrow$fcc transition state ($\alpha\approx 0.54$), fcc ($\alpha=1, \gamma_2=\sqrt{2}$) and bcc ($\alpha=1,  \gamma_2=1$) structures.}
    \label{fig:hypersurfaces}
\end{figure}

The $\Delta E_{\text{coh}}(\alpha,\gamma_2)=E_{\text{coh}}(0,\sqrt{\frac{8}{3}})-E_{\text{coh}}(\alpha,\gamma_2)$ hypersurface for the (12-6)-LJ  potential (taking the hcp structure as the reference) is shown in Figure \ref{fig:hypersurfaces}. It is clear that there is no direct MEP from hcp to bcc; the MEP starting at hcp leads directly to the fcc structure as predicted from the simple KHS model. Further calculations reveal that neither of the more general $(n,m)$-LJ potentials have a direct MEP from hcp to bcc \cite{schwerdtfeger2024}. Starting from bcc towards the hcp structure, by changing the parameter $\alpha$ and optimizing all others, ends in steep ascend energetically. As discussed before \cite{burrows2021b}, the bcc structure at $\alpha=1$ and $\gamma_2=1$ is a maximum for the (12,6)-LJ potential. Concerning the transition state for the hcp$\rightarrow$fcc path, we locate it at $\alpha^{\#}=0.50036605$ with an activation energy of $\Delta E_\text{coh}^{\#}=0.41488952$ relative to the hcp structure. The volume at the transition state is 1.32709616 and below the KHS limit as we expect for soft penetrating spheres.

 For a real quantum system, we consider metallic lithium analyzed at the density functional theory (DFT) level. A detailed account of the lithium phase diagram has been given by Guillaume et al. \cite{Guillaume2011} and Ackland et al. \cite{Ackland2017} Each point on the $(\alpha,\gamma_{2})$-energy hypersurface, as shown in Figure \ref{fig:hypersurfaces},  corresponds to an optimization of $a$ and $\beta_{2}$ at fixed values of $\alpha$ and $\gamma_{2}$, together with $\gamma_{1}=1$, $\beta_{1}=1$ and $\beta_{3}=\gamma_{2}$. The algorithm employed in the crystal structure optimization is the Newton-Raphson with the gradient and Hessian matrix calculated numerically through finite differences and ensuring an energy threshold for convergence of $10^{-6}$ eV in the electronic energy. The single-point calculations were performed with density functional theory using the Perdew-Burke-Ernzerhof (PBE) exchange-correlation functional \cite{Perdew1996} coupled to the dispersion corrections by means of the DFT-D3 approach including the Becke-Johnson damping function \cite{Grimme2010,Grimme2011}, as implemented in the VASP 6.4.0 version \cite{VASP1,VASP2,VASP3,VASP4,VASP5}. The atomic cores are described by the Projector Augmented Wave (PAW) method \cite{Blochl1994PAW}, and the electronic minimization was done using the tetrahedron method \cite{Blochl1994BZInt} with Bl\"ochl correction with an energy width of $0.1$ eV. The $k$-point grid was set using the keyword KSPACING=0.07 centered at the $\Gamma$ point and an energy cut-off of 500 eV. The cohesive energies are obtained by comparing the energy of the bulk and the isolated atom, the latter calculated from a large orthorhombic unit cell of $14\times 14.001\times 14.002$ \AA. Scanning the hypersurface and optimizing the lattice parameters is computationally very demanding.

\begin{center}
\begin{table}[hbtp!] 
\setlength{\tabcolsep}{1.3pt}
\begin{tabular}{|c|c|c|c|c|c|} 
\hline
Prop.        & hcp         & fcc          & bcc        & $\text{TS}_{1}$ & $\text{TS}_{2}$\\ \hline 
$E_{\text{coh}}$& $172.2216$  & $172.2515$   & $172.0344$ & $171.1907$      & $172.0212$\\
$\alpha$        & $0.000$     & $1.000$      & $1.000$    & $0.541$         & $1.000$ \\
$a$             & $2.962$     & $2.963$      & $3.325$    & $2.9079$        & $3.230$ \\
$\gamma_2$      & $1.632$     & $1.411$      & $1.000$    & $1.599$         & $1.095$ \\
$\beta_2$       & $0.577$     & $1.000$      & $1.000$    & $0.820$         & $1.000$ \\
$V$             & $36.731$    & $36.708$     & $36.757$   & $38.268$        & $36.900$ \\
\hline
\end{tabular}
\caption{Lattice parameters and volume $V$ for the hcp, fcc and bcc structures of lithium together with the transition states along the Burgers-Bain transformation path. The transition state of the hcp$\leftrightarrow$fcc transformation is denoted as $\text{TS}_{1}$, whereas for the fcc$\leftrightarrow$bcc is denoted by $\text{TS}_{2}$.}
\label{tab:lithium-lattice-param}
\end{table}
\end{center}

Figure \ref{fig:hypersurfaces} shows that the topology of the lithium cohesive energy surface is quite similar to the one obtained for the $(12,6)$-LJ potential, with the deepest minima located at the hcp and the fcc structures, fcc being the ground state, see Table \ref{tab:lithium-lattice-param}. This is in agreement with a detailed phase diagram study of Ackland et al. \cite{Ackland2017} At standard conditions, the lithium metal is found in the bcc structure with a measured cohesive energy of $158$ kJ/mol \cite{Kittel2004}. The difference from the experimental cohesive energy comes from temperature effects, zero-point vibrational contributions (3.971 kJ/mol for the fcc structure using the PBE functional \cite{RoblesNavarro2024}) and the density functional approximation applied. We expect that this error is mostly compensated when taking energy differences along the transformation path \cite{RoblesNavarro2024}. The bcc structure of lithium is shown to be a very shallow minimum, in agreement with previous work \cite{Jerabek2022,RoblesNavarro2024}. The path connecting hcp to the cuboidal structures leads directly to the fcc structure and has a transition state at $\alpha^{\#}=0.541$ with an energy difference from hcp of $\Delta E_\text{coh}^{\#}=1.031$ kJ/mol. The transformation from fcc to bcc occurs through a second transition state at $\gamma_{2}=1.095$ with an energy difference of $\Delta E_\text{coh}^{\#}=0.230$ kJ/mol from the fcc phase. Noticeably, the hcp$\rightarrow$fcc transition state lies energetically above the fcc$\rightarrow$bcc Bain path. The maximum volume increase for the hcp$\rightarrow$fcc transition is $4.18 \%$ compared to the hcp structure (see Table \ref{tab:lithium-lattice-param}), and is less compared to the hard-sphere limit of $6.07\%$ $(3/2\sqrt{2})$ as we expect.

In summary, for the unit cell, lattice parameters and three different interaction models applied we do not observe a direct minimum energy path from hcp to bcc as originally suggested by Burgers \cite{Burgers1934} and others \cite{Li1999,Natarajan2019}. The transition states on the $(\alpha,\gamma_2)$-hypersurface MEP for lithium comes at rather low energies of about 1 kJ/mol. Classical molecular dynamics simulations at higher temperatures will explore the whole region of the parameter space outside the MEP \cite{Bingxi2017}, and from such simulations it is therefore often difficult to get a detailed picture of the phase transition compared to directly mapping out the MEP as presented here. For solid-state systems such as barium,\cite{Chen1988} or iron\cite{Ekman1998}, where the bcc phase is significantly stabilized over the fcc or hcp phase \cite{Grimvall2012}, the topology of the hypersurface is expected to change, which is the subject of our future investigations.\\

\section*{Acknowledgments}
PS is grateful to Prof. Paul Indelicato (Kastler lab, Paris) for a visiting professorship at Sorbonne.

\bibliography{references}

\providecommand{\latin}[1]{#1}
\makeatletter
\providecommand{\doi}
  {\begingroup\let\do\@makeother\dospecials
  \catcode`\{=1 \catcode`\}=2 \doi@aux}
\providecommand{\doi@aux}[1]{\endgroup\texttt{#1}}
\makeatother
\providecommand*\mcitethebibliography{\thebibliography}
\csname @ifundefined\endcsname{endmcitethebibliography}
  {\let\endmcitethebibliography\endthebibliography}{}
\begin{mcitethebibliography}{52}
\providecommand*\natexlab[1]{#1}
\providecommand*\mciteSetBstSublistMode[1]{}
\providecommand*\mciteSetBstMaxWidthForm[2]{}
\providecommand*\mciteBstWouldAddEndPuncttrue
  {\def\EndOfBibitem{\unskip.}}
\providecommand*\mciteBstWouldAddEndPunctfalse
  {\let\EndOfBibitem\relax}
\providecommand*\mciteSetBstMidEndSepPunct[3]{}
\providecommand*\mciteSetBstSublistLabelBeginEnd[3]{}
\providecommand*\EndOfBibitem{}
\mciteSetBstSublistMode{f}
\mciteSetBstMaxWidthForm{subitem}{(\alph{mcitesubitemcount})}
\mciteSetBstSublistLabelBeginEnd
  {\mcitemaxwidthsubitemform\space}
  {\relax}
  {\relax}

\bibitem[{Liu, Xueyan} \latin{et~al.}(2018){Liu, Xueyan}, {Li, Hongwei}, and
  {Zhan, Mei}]{Liu2018}
{Liu, Xueyan},; {Li, Hongwei},; {Zhan, Mei}, A review on the modeling and
  simulations of solid-state diffusional phase transformations in metals and
  alloys. \emph{Manufacturing Rev.} \textbf{2018}, \emph{5}, 10\relax
\mciteBstWouldAddEndPuncttrue
\mciteSetBstMidEndSepPunct{\mcitedefaultmidpunct}
{\mcitedefaultendpunct}{\mcitedefaultseppunct}\relax
\EndOfBibitem
\bibitem[Hansen and Verlet(1969)Hansen, and Verlet]{Verlet1969}
Hansen,~J.-P.; Verlet,~L. {Phase Transitions of the Lennard-Jones System}.
  \emph{Phys. Rev.} \textbf{1969}, \emph{184}, 151--161\relax
\mciteBstWouldAddEndPuncttrue
\mciteSetBstMidEndSepPunct{\mcitedefaultmidpunct}
{\mcitedefaultendpunct}{\mcitedefaultseppunct}\relax
\EndOfBibitem
\bibitem[Chang \latin{et~al.}(2004)Chang, Chen, Zhang, Yan, Xie, Schmid-Fetzer,
  and Oates]{CHANG2004}
Chang,~Y.; Chen,~S.; Zhang,~F.; Yan,~X.; Xie,~F.; Schmid-Fetzer,~R.; Oates,~W.
  Phase diagram calculation: past, present and future. \emph{Prog. Mat. Sci.}
  \textbf{2004}, \emph{49}, 313--345\relax
\mciteBstWouldAddEndPuncttrue
\mciteSetBstMidEndSepPunct{\mcitedefaultmidpunct}
{\mcitedefaultendpunct}{\mcitedefaultseppunct}\relax
\EndOfBibitem
\bibitem[Schlegel(2003)]{Schlegel2003}
Schlegel,~H.~B. Exploring potential energy surfaces for chemical reactions: An
  overview of some practical methods. \emph{J. Comput. Chem.} \textbf{2003},
  \emph{24}, 1514--1527\relax
\mciteBstWouldAddEndPuncttrue
\mciteSetBstMidEndSepPunct{\mcitedefaultmidpunct}
{\mcitedefaultendpunct}{\mcitedefaultseppunct}\relax
\EndOfBibitem
\bibitem[Bartlett and Musia\l{}(2007)Bartlett, and Musia\l{}]{Bartlett2007}
Bartlett,~R.~J.; Musia\l{},~M. Coupled-cluster theory in quantum chemistry.
  \emph{Rev. Mod. Phys.} \textbf{2007}, \emph{79}, 291--352\relax
\mciteBstWouldAddEndPuncttrue
\mciteSetBstMidEndSepPunct{\mcitedefaultmidpunct}
{\mcitedefaultendpunct}{\mcitedefaultseppunct}\relax
\EndOfBibitem
\bibitem[Caspersen and Carter(2005)Caspersen, and Carter]{Carter2005}
Caspersen,~K.~J.; Carter,~E.~A. Finding transition states for crystalline
  solid--solid phase transformations. \emph{Proc. Nat. Acad. Sci.}
  \textbf{2005}, \emph{102}, 6738--6743\relax
\mciteBstWouldAddEndPuncttrue
\mciteSetBstMidEndSepPunct{\mcitedefaultmidpunct}
{\mcitedefaultendpunct}{\mcitedefaultseppunct}\relax
\EndOfBibitem
\bibitem[Travesset(2014)]{Travesset2014}
Travesset,~A. Phase diagram of power law and Lennard-Jones systems: Crystal
  phases. \emph{J. Chem. Phys.} \textbf{2014}, \emph{141}, 164501\relax
\mciteBstWouldAddEndPuncttrue
\mciteSetBstMidEndSepPunct{\mcitedefaultmidpunct}
{\mcitedefaultendpunct}{\mcitedefaultseppunct}\relax
\EndOfBibitem
\bibitem[de~Wijs \latin{et~al.}(1998)de~Wijs, Kresse, and Gillan]{Kresse1998}
de~Wijs,~G.~A.; Kresse,~G.; Gillan,~M.~J. First-order phase transitions by
  first-principles free-energy calculations: The melting of {A}l. \emph{Phys.
  Rev. B} \textbf{1998}, \emph{57}, 8223--8234\relax
\mciteBstWouldAddEndPuncttrue
\mciteSetBstMidEndSepPunct{\mcitedefaultmidpunct}
{\mcitedefaultendpunct}{\mcitedefaultseppunct}\relax
\EndOfBibitem
\bibitem[Jinnouchi \latin{et~al.}(2019)Jinnouchi, Lahnsteiner, Karsai, Kresse,
  and Bokdam]{Kresse2019}
Jinnouchi,~R.; Lahnsteiner,~J.; Karsai,~F.; Kresse,~G.; Bokdam,~M. {Phase
  Transitions of Hybrid Perovskites Simulated by Machine-Learning Force Fields
  Trained on the Fly with Bayesian Inference}. \emph{Phys. Rev. Lett.}
  \textbf{2019}, \emph{122}, 225701\relax
\mciteBstWouldAddEndPuncttrue
\mciteSetBstMidEndSepPunct{\mcitedefaultmidpunct}
{\mcitedefaultendpunct}{\mcitedefaultseppunct}\relax
\EndOfBibitem
\bibitem[Liu \latin{et~al.}(2021)Liu, Verdi, Karsai, and Kresse]{Kresse2021}
Liu,~P.; Verdi,~C.; Karsai,~F.; Kresse,~G.
  $\ensuremath{\alpha}\text{\ensuremath{-}}\ensuremath{\beta}$ phase transition
  of zirconium predicted by on-the-fly machine-learned force field. \emph{Phys.
  Rev. Mater.} \textbf{2021}, \emph{5}, 053804\relax
\mciteBstWouldAddEndPuncttrue
\mciteSetBstMidEndSepPunct{\mcitedefaultmidpunct}
{\mcitedefaultendpunct}{\mcitedefaultseppunct}\relax
\EndOfBibitem
\bibitem[Grimvall \latin{et~al.}(2012)Grimvall, Magyari-K\"ope, Ozoli\ifmmode
  \mbox{\c{n}}\else \c{n}\fi{}\ifmmode~\check{s}\else \v{s}\fi{}, and
  Persson]{Grimvall2012}
Grimvall,~G.; Magyari-K\"ope,~B.; Ozoli\ifmmode \mbox{\c{n}}\else
  \c{n}\fi{}\ifmmode~\check{s}\else \v{s}\fi{},~V.; Persson,~K.~A. Lattice
  instabilities in metallic elements. \emph{Rev. Mod. Phys.} \textbf{2012},
  \emph{84}, 945--986\relax
\mciteBstWouldAddEndPuncttrue
\mciteSetBstMidEndSepPunct{\mcitedefaultmidpunct}
{\mcitedefaultendpunct}{\mcitedefaultseppunct}\relax
\EndOfBibitem
\bibitem[Nishiyama(2012)]{nishiyama2012}
Nishiyama,~Z. \emph{Martensitic transformation}; Elsevier, Amsterdam,
  2012\relax
\mciteBstWouldAddEndPuncttrue
\mciteSetBstMidEndSepPunct{\mcitedefaultmidpunct}
{\mcitedefaultendpunct}{\mcitedefaultseppunct}\relax
\EndOfBibitem
\bibitem[Lobodyuk and Estrin(2014)Lobodyuk, and Estrin]{lobodyuk2014}
Lobodyuk,~V.; Estrin,~E. \emph{Martensitic Transformations}; Cambridge
  International Science Publishing, 2014\relax
\mciteBstWouldAddEndPuncttrue
\mciteSetBstMidEndSepPunct{\mcitedefaultmidpunct}
{\mcitedefaultendpunct}{\mcitedefaultseppunct}\relax
\EndOfBibitem
\bibitem[Venables(1962)]{Venables1962}
Venables,~J.~A. The martensite transformation in stainless steel. \emph{Phil.
  Mag.} \textbf{1962}, \emph{7}, 35--44\relax
\mciteBstWouldAddEndPuncttrue
\mciteSetBstMidEndSepPunct{\mcitedefaultmidpunct}
{\mcitedefaultendpunct}{\mcitedefaultseppunct}\relax
\EndOfBibitem
\bibitem[Bain(1924)]{Bain1924}
Bain,~E. The Nature of Martensite. \emph{Trans. Am. Inst. Min. Metall. Eng.}
  \textbf{1924}, \emph{70}, 25--46\relax
\mciteBstWouldAddEndPuncttrue
\mciteSetBstMidEndSepPunct{\mcitedefaultmidpunct}
{\mcitedefaultendpunct}{\mcitedefaultseppunct}\relax
\EndOfBibitem
\bibitem[Burgers(1934)]{Burgers1934}
Burgers,~W. On the process of transition of the cubic-body-centered
  modification into the hexagonal-close-packed modification of zirconium.
  \emph{Physica} \textbf{1934}, \emph{1}, 561--586\relax
\mciteBstWouldAddEndPuncttrue
\mciteSetBstMidEndSepPunct{\mcitedefaultmidpunct}
{\mcitedefaultendpunct}{\mcitedefaultseppunct}\relax
\EndOfBibitem
\bibitem[Cayron(2015)]{Cayron2015}
Cayron,~C. Continuous atomic displacements and lattice distortion during
  fcc–bcc martensitic transformation. \emph{Acta Mat.} \textbf{2015},
  \emph{96}, 189--202\relax
\mciteBstWouldAddEndPuncttrue
\mciteSetBstMidEndSepPunct{\mcitedefaultmidpunct}
{\mcitedefaultendpunct}{\mcitedefaultseppunct}\relax
\EndOfBibitem
\bibitem[Olson and Cohen(1976)Olson, and Cohen]{Olson1976}
Olson,~G.~B.; Cohen,~M. A general mechanism of martensitic nucleation: Part I.
  General concepts and the fcc →hcp transformation. \emph{Metal. Trans. A}
  \textbf{1976}, \emph{7}, 1897--1904\relax
\mciteBstWouldAddEndPuncttrue
\mciteSetBstMidEndSepPunct{\mcitedefaultmidpunct}
{\mcitedefaultendpunct}{\mcitedefaultseppunct}\relax
\EndOfBibitem
\bibitem[Bruinsma and Zangwill(1985)Bruinsma, and Zangwill]{Bruinsma1985}
Bruinsma,~R.; Zangwill,~A. Theory of the hcp-fcc transition in metals.
  \emph{Phys. Rev. Lett.} \textbf{1985}, \emph{55}, 214--217\relax
\mciteBstWouldAddEndPuncttrue
\mciteSetBstMidEndSepPunct{\mcitedefaultmidpunct}
{\mcitedefaultendpunct}{\mcitedefaultseppunct}\relax
\EndOfBibitem
\bibitem[Akahama \latin{et~al.}(2006)Akahama, Nishimura, Kinoshita, Kawamura,
  and Ohishi]{Akahama2006}
Akahama,~Y.; Nishimura,~M.; Kinoshita,~K.; Kawamura,~H.; Ohishi,~Y. Evidence of
  a fcc-hcp Transition in Aluminum at Multimegabar Pressure. \emph{Phys. Rev.
  Lett.} \textbf{2006}, \emph{96}, 045505\relax
\mciteBstWouldAddEndPuncttrue
\mciteSetBstMidEndSepPunct{\mcitedefaultmidpunct}
{\mcitedefaultendpunct}{\mcitedefaultseppunct}\relax
\EndOfBibitem
\bibitem[Lu \latin{et~al.}(2014)Lu, Zhu, Lu, and Wang]{Lu_2014}
Lu,~Z.; Zhu,~W.; Lu,~T.; Wang,~W. Does the fcc phase exist in the Fe bcc--hcp
  transition? {A} conclusion from first-principles studies. \emph{Model. Simul.
  Mater. Sci. Eng.} \textbf{2014}, \emph{22}, 025007\relax
\mciteBstWouldAddEndPuncttrue
\mciteSetBstMidEndSepPunct{\mcitedefaultmidpunct}
{\mcitedefaultendpunct}{\mcitedefaultseppunct}\relax
\EndOfBibitem
\bibitem[Baxter(1968)]{Baxter1968}
Baxter,~R.~J. {{Percus–Yevick Equation for Hard Spheres with Surface
  Adhesion}}. \emph{J. Chem. Phys.} \textbf{1968}, \emph{49}, 2770--2774\relax
\mciteBstWouldAddEndPuncttrue
\mciteSetBstMidEndSepPunct{\mcitedefaultmidpunct}
{\mcitedefaultendpunct}{\mcitedefaultseppunct}\relax
\EndOfBibitem
\bibitem[Schwerdtfeger and Wales(2024)Schwerdtfeger, and Wales]{Wales2024}
Schwerdtfeger,~P.; Wales,~D.~J. {100 Years of the Lennard-Jones Potential}.
  \emph{J. Chem. Theory Comput.} \textbf{2024}, \emph{20}, 3379--3405\relax
\mciteBstWouldAddEndPuncttrue
\mciteSetBstMidEndSepPunct{\mcitedefaultmidpunct}
{\mcitedefaultendpunct}{\mcitedefaultseppunct}\relax
\EndOfBibitem
\bibitem[Gross and Dreizler(2013)Gross, and Dreizler]{gross2013}
Gross,~E.~K.; Dreizler,~R.~M. \emph{Density functional theory}; Springer
  Science \& Business Media, 2013; Vol. 337\relax
\mciteBstWouldAddEndPuncttrue
\mciteSetBstMidEndSepPunct{\mcitedefaultmidpunct}
{\mcitedefaultendpunct}{\mcitedefaultseppunct}\relax
\EndOfBibitem
\bibitem[Straub and Wallace(1971)Straub, and Wallace]{Straub1971}
Straub,~G.~K.; Wallace,~D.~C. Study of the Martensitic Phase Transition in
  Sodium. \emph{Phys. Rev. B} \textbf{1971}, \emph{3}, 1234--1239\relax
\mciteBstWouldAddEndPuncttrue
\mciteSetBstMidEndSepPunct{\mcitedefaultmidpunct}
{\mcitedefaultendpunct}{\mcitedefaultseppunct}\relax
\EndOfBibitem
\bibitem[Li and Wang(1999)Li, and Wang]{Li1999}
Li,~W.; Wang,~T. Theoretical investigation of epitaxial deformation and the
  hcp-bcc transition of alkali metals. \emph{Phys. Rev. B} \textbf{1999},
  \emph{60}, 11954--11962\relax
\mciteBstWouldAddEndPuncttrue
\mciteSetBstMidEndSepPunct{\mcitedefaultmidpunct}
{\mcitedefaultendpunct}{\mcitedefaultseppunct}\relax
\EndOfBibitem
\bibitem[Raju~Natarajan and Van~der Ven(2019)Raju~Natarajan, and Van~der
  Ven]{Natarajan2019}
Raju~Natarajan,~A.; Van~der Ven,~A. Toward an Understanding of Deformation
  Mechanisms in Metallic Lithium and Sodium from First-Principles. \emph{Chem.
  Mater.} \textbf{2019}, \emph{31}, 8222--8229\relax
\mciteBstWouldAddEndPuncttrue
\mciteSetBstMidEndSepPunct{\mcitedefaultmidpunct}
{\mcitedefaultendpunct}{\mcitedefaultseppunct}\relax
\EndOfBibitem
\bibitem[Chen \latin{et~al.}(1988)Chen, Ho, and Harmon]{Chen1988}
Chen,~Y.; Ho,~K.~M.; Harmon,~B.~N. First-principles study of the
  pressure-induced bcc-hcp transition in Ba. \emph{Phys. Rev. B} \textbf{1988},
  \emph{37}, 283--288\relax
\mciteBstWouldAddEndPuncttrue
\mciteSetBstMidEndSepPunct{\mcitedefaultmidpunct}
{\mcitedefaultendpunct}{\mcitedefaultseppunct}\relax
\EndOfBibitem
\bibitem[Sutherland(1893)]{Sutherland1893}
Sutherland,~W. LII. The viscosity of gases and molecular force. \emph{London
  Edinburgh Philos. Mag. \& J. Sci.} \textbf{1893}, \emph{36}, 507--531\relax
\mciteBstWouldAddEndPuncttrue
\mciteSetBstMidEndSepPunct{\mcitedefaultmidpunct}
{\mcitedefaultendpunct}{\mcitedefaultseppunct}\relax
\EndOfBibitem
\bibitem[Burrows \latin{et~al.}(2020)Burrows, Cooper, Pahl, and
  Schwerdtfeger]{burrows-2020}
Burrows,~A.; Cooper,~S.; Pahl,~E.; Schwerdtfeger,~P. Analytical methods for
  fast converging lattice sums for cubic and hexagonal close-packed structures.
  \emph{J. Math. Phys.} \textbf{2020}, \emph{61}, 123503\relax
\mciteBstWouldAddEndPuncttrue
\mciteSetBstMidEndSepPunct{\mcitedefaultmidpunct}
{\mcitedefaultendpunct}{\mcitedefaultseppunct}\relax
\EndOfBibitem
\bibitem[Burrows \latin{et~al.}(2021)Burrows, Cooper, and
  Schwerdtfeger]{burrows2021b}
Burrows,~A.; Cooper,~S.; Schwerdtfeger,~P. Instability of the body-centered
  cubic lattice within the sticky hard sphere and {L}ennard-{J}ones model
  obtained from exact lattice summations. \emph{Phys. Rev. E} \textbf{2021},
  \emph{104}, 035306\relax
\mciteBstWouldAddEndPuncttrue
\mciteSetBstMidEndSepPunct{\mcitedefaultmidpunct}
{\mcitedefaultendpunct}{\mcitedefaultseppunct}\relax
\EndOfBibitem
\bibitem[Schwerdtfeger \latin{et~al.}(2006)Schwerdtfeger, Gaston, Krawczyk,
  Tonner, and Moyano]{Schwerdtfeger-2006}
Schwerdtfeger,~P.; Gaston,~N.; Krawczyk,~R.~P.; Tonner,~R.; Moyano,~G.~E.
  {Extension of the Lennard-Jones potential: Theoretical investigations into
  rare-gas clusters and crystal lattices of He, Ne, Ar, and Kr using many-body
  interaction expansions}. \emph{Phys. Rev. B} \textbf{2006}, \emph{73},
  064112\relax
\mciteBstWouldAddEndPuncttrue
\mciteSetBstMidEndSepPunct{\mcitedefaultmidpunct}
{\mcitedefaultendpunct}{\mcitedefaultseppunct}\relax
\EndOfBibitem
\bibitem[H{\"a}mmerlin and Hoffmann(2012)H{\"a}mmerlin, and
  Hoffmann]{hammerlin2012numerical}
H{\"a}mmerlin,~G.; Hoffmann,~K.-H. \emph{Numerical mathematics}; Springer
  Science \& Business Media, Berlin, 2012\relax
\mciteBstWouldAddEndPuncttrue
\mciteSetBstMidEndSepPunct{\mcitedefaultmidpunct}
{\mcitedefaultendpunct}{\mcitedefaultseppunct}\relax
\EndOfBibitem
\bibitem[Schwerdtfeger \latin{et~al.}(2024)Schwerdtfeger, Cooper, Smits, and
  Robles-Navarro]{schwerdtfeger2024}
Schwerdtfeger,~P.; Cooper,~S.; Smits,~O.; Robles-Navarro,~A. {Connecting the
  hexagonal closed packed structure with the cuboidal lattices: A Burgers-Bain
  type martensitic transformation for a Lennard-Jones solid derived from exact
  lattice summations}. 2024; \url{https://arxiv.org/abs/2406.09635}\relax
\mciteBstWouldAddEndPuncttrue
\mciteSetBstMidEndSepPunct{\mcitedefaultmidpunct}
{\mcitedefaultendpunct}{\mcitedefaultseppunct}\relax
\EndOfBibitem
\bibitem[Guillaume \latin{et~al.}(2011)Guillaume, Gregoryanz, Degtyareva,
  McMahon, Hanfland, Evans, Guthrie, Sinogeikin, and Mao]{Guillaume2011}
Guillaume,~C.~L.; Gregoryanz,~E.; Degtyareva,~O.; McMahon,~M.~I.; Hanfland,~M.;
  Evans,~S.; Guthrie,~M.; Sinogeikin,~S.~V.; Mao,~H.-K. Cold melting and solid
  structures of dense lithium. \emph{Nature Physics} \textbf{2011}, \emph{7},
  211--214\relax
\mciteBstWouldAddEndPuncttrue
\mciteSetBstMidEndSepPunct{\mcitedefaultmidpunct}
{\mcitedefaultendpunct}{\mcitedefaultseppunct}\relax
\EndOfBibitem
\bibitem[Ackland \latin{et~al.}(2017)Ackland, Dunuwille, Martinez-Canales, Loa,
  Zhang, Sinogeikin, Cai, and Deemyad]{Ackland2017}
Ackland,~G.~J.; Dunuwille,~M.; Martinez-Canales,~M.; Loa,~I.; Zhang,~R.;
  Sinogeikin,~S.; Cai,~W.; Deemyad,~S. Quantum and isotope effects in lithium
  metal. \emph{Science} \textbf{2017}, \emph{356}, 1254--1259\relax
\mciteBstWouldAddEndPuncttrue
\mciteSetBstMidEndSepPunct{\mcitedefaultmidpunct}
{\mcitedefaultendpunct}{\mcitedefaultseppunct}\relax
\EndOfBibitem
\bibitem[Perdew \latin{et~al.}(1996)Perdew, Burke, and Ernzerhof]{Perdew1996}
Perdew,~J.~P.; Burke,~K.; Ernzerhof,~M. Generalized Gradient Approximation Made
  Simple. \emph{Phys. Rev. Lett.} \textbf{1996}, \emph{77}, 3865--3868\relax
\mciteBstWouldAddEndPuncttrue
\mciteSetBstMidEndSepPunct{\mcitedefaultmidpunct}
{\mcitedefaultendpunct}{\mcitedefaultseppunct}\relax
\EndOfBibitem
\bibitem[Grimme \latin{et~al.}(2010)Grimme, Antony, Ehrlich, and
  Krieg]{Grimme2010}
Grimme,~S.; Antony,~J.; Ehrlich,~S.; Krieg,~H. {A consistent and accurate ab
  initio parametrization of density functional dispersion correction (DFT-D)
  for the 94 elements {H-Pu}}. \emph{J. Chem. Phys.} \textbf{2010}, \emph{132},
  154104\relax
\mciteBstWouldAddEndPuncttrue
\mciteSetBstMidEndSepPunct{\mcitedefaultmidpunct}
{\mcitedefaultendpunct}{\mcitedefaultseppunct}\relax
\EndOfBibitem
\bibitem[Grimme \latin{et~al.}(2011)Grimme, Ehrlich, and Goerigk]{Grimme2011}
Grimme,~S.; Ehrlich,~S.; Goerigk,~L. Effect of the damping function in
  dispersion corrected density functional theory. \emph{J. Comput. Chem.}
  \textbf{2011}, \emph{32}, 1456--1465\relax
\mciteBstWouldAddEndPuncttrue
\mciteSetBstMidEndSepPunct{\mcitedefaultmidpunct}
{\mcitedefaultendpunct}{\mcitedefaultseppunct}\relax
\EndOfBibitem
\bibitem[Kresse and Hafner(1993)Kresse, and Hafner]{VASP1}
Kresse,~G.; Hafner,~J. Ab initio molecular dynamics for liquid metals.
  \emph{Phys. Rev. B} \textbf{1993}, \emph{47}, 558--561\relax
\mciteBstWouldAddEndPuncttrue
\mciteSetBstMidEndSepPunct{\mcitedefaultmidpunct}
{\mcitedefaultendpunct}{\mcitedefaultseppunct}\relax
\EndOfBibitem
\bibitem[Kresse and Hafner(1994)Kresse, and Hafner]{VASP2}
Kresse,~G.; Hafner,~J. Ab initio molecular-dynamics simulation of the
  liquid-metal--amorphous-semiconductor transition in germanium. \emph{Phys.
  Rev. B} \textbf{1994}, \emph{49}, 14251--14269\relax
\mciteBstWouldAddEndPuncttrue
\mciteSetBstMidEndSepPunct{\mcitedefaultmidpunct}
{\mcitedefaultendpunct}{\mcitedefaultseppunct}\relax
\EndOfBibitem
\bibitem[Kresse and Furthmüller(1996)Kresse, and Furthmüller]{VASP3}
Kresse,~G.; Furthmüller,~J. Efficiency of ab-initio total energy calculations
  for metals and semiconductors using a plane-wave basis set. \emph{Comput.
  Mat. Sci.} \textbf{1996}, \emph{6}, 15--50\relax
\mciteBstWouldAddEndPuncttrue
\mciteSetBstMidEndSepPunct{\mcitedefaultmidpunct}
{\mcitedefaultendpunct}{\mcitedefaultseppunct}\relax
\EndOfBibitem
\bibitem[Kresse and Furthm\"uller(1996)Kresse, and Furthm\"uller]{VASP4}
Kresse,~G.; Furthm\"uller,~J. Efficient iterative schemes for ab initio
  total-energy calculations using a plane-wave basis set. \emph{Phys. Rev. B}
  \textbf{1996}, \emph{54}, 11169--11186\relax
\mciteBstWouldAddEndPuncttrue
\mciteSetBstMidEndSepPunct{\mcitedefaultmidpunct}
{\mcitedefaultendpunct}{\mcitedefaultseppunct}\relax
\EndOfBibitem
\bibitem[Kresse and Joubert(1999)Kresse, and Joubert]{VASP5}
Kresse,~G.; Joubert,~D. From ultrasoft pseudopotentials to the projector
  augmented-wave method. \emph{Phys. Rev. B} \textbf{1999}, \emph{59},
  1758--1775\relax
\mciteBstWouldAddEndPuncttrue
\mciteSetBstMidEndSepPunct{\mcitedefaultmidpunct}
{\mcitedefaultendpunct}{\mcitedefaultseppunct}\relax
\EndOfBibitem
\bibitem[Bl\"ochl(1994)]{Blochl1994PAW}
Bl\"ochl,~P.~E. Projector augmented-wave method. \emph{Phys. Rev. B}
  \textbf{1994}, \emph{50}, 17953--17979\relax
\mciteBstWouldAddEndPuncttrue
\mciteSetBstMidEndSepPunct{\mcitedefaultmidpunct}
{\mcitedefaultendpunct}{\mcitedefaultseppunct}\relax
\EndOfBibitem
\bibitem[Bl\"ochl \latin{et~al.}(1994)Bl\"ochl, Jepsen, and
  Andersen]{Blochl1994BZInt}
Bl\"ochl,~P.~E.; Jepsen,~O.; Andersen,~O.~K. Improved tetrahedron method for
  {B}rillouin-zone integrations. \emph{Phys. Rev. B} \textbf{1994}, \emph{49},
  16223--16233\relax
\mciteBstWouldAddEndPuncttrue
\mciteSetBstMidEndSepPunct{\mcitedefaultmidpunct}
{\mcitedefaultendpunct}{\mcitedefaultseppunct}\relax
\EndOfBibitem
\bibitem[Kittel(2004)]{Kittel2004}
Kittel,~C. \emph{Introduction to Solid State Physics}; John Wiley \& Sons,
  Inc.: New York, 2004\relax
\mciteBstWouldAddEndPuncttrue
\mciteSetBstMidEndSepPunct{\mcitedefaultmidpunct}
{\mcitedefaultendpunct}{\mcitedefaultseppunct}\relax
\EndOfBibitem
\bibitem[Robles-Navarro \latin{et~al.}(2024)Robles-Navarro, Jerabek, and
  Schwerdtfeger]{RoblesNavarro2024}
Robles-Navarro,~A.; Jerabek,~P.; Schwerdtfeger,~P. Tipping the Balance Between
  the bcc and fcc Phase Within the Alkali and Coinage Metal Groups.
  \emph{Angew. Chem. Int. Ed.} \textbf{2024}, \emph{63}, e202313679\relax
\mciteBstWouldAddEndPuncttrue
\mciteSetBstMidEndSepPunct{\mcitedefaultmidpunct}
{\mcitedefaultendpunct}{\mcitedefaultseppunct}\relax
\EndOfBibitem
\bibitem[Jerabek \latin{et~al.}(2022)Jerabek, Burrows, and
  Schwerdtfeger]{Jerabek2022}
Jerabek,~P.; Burrows,~A.; Schwerdtfeger,~P. Solving a problem with a single
  parameter: a smooth bcc to fcc phase transition for metallic lithium.
  \emph{Chem. Commun.} \textbf{2022}, \emph{58}, 13369--13372\relax
\mciteBstWouldAddEndPuncttrue
\mciteSetBstMidEndSepPunct{\mcitedefaultmidpunct}
{\mcitedefaultendpunct}{\mcitedefaultseppunct}\relax
\EndOfBibitem
\bibitem[Li \latin{et~al.}(2017)Li, Qian, Oganov, Boulfelfel, and
  Faller]{Bingxi2017}
Li,~B.; Qian,~G.; Oganov,~A.~R.; Boulfelfel,~S.~E.; Faller,~R. Mechanism of the
  fcc-to-hcp phase transformation in solid Ar. \emph{J. Chem. Phys.}
  \textbf{2017}, \emph{146}, 214502\relax
\mciteBstWouldAddEndPuncttrue
\mciteSetBstMidEndSepPunct{\mcitedefaultmidpunct}
{\mcitedefaultendpunct}{\mcitedefaultseppunct}\relax
\EndOfBibitem
\bibitem[Ekman \latin{et~al.}(1998)Ekman, Sadigh, Einarsdotter, and
  Blaha]{Ekman1998}
Ekman,~M.; Sadigh,~B.; Einarsdotter,~K.; Blaha,~P. Ab initio study of the
  martensitic bcc-hcp transformation in iron. \emph{Phys. Rev. B}
  \textbf{1998}, \emph{58}, 5296--5304\relax
\mciteBstWouldAddEndPuncttrue
\mciteSetBstMidEndSepPunct{\mcitedefaultmidpunct}
{\mcitedefaultendpunct}{\mcitedefaultseppunct}\relax
\EndOfBibitem
\end{mcitethebibliography}

\end{document}